\documentclass[pre,twocolumn,showpacs,preprintnumbers,amsmath,amssymb]{revtex4} 
\usepackage{amsmath}
\usepackage{graphicx}
\usepackage{dcolumn}
\usepackage{bm}

\def\max{\mathrm{max}}

\begin{document}
\title{Solvable senescence model with positive mutations}
\author{J.~B.~Coe$^1$, Y.~Mao$^{1,2}$ and M.~E.~Cates$^3$}
\affiliation{$^1$Cavendish Laboratory, Madingley Road, Cambridge CB3 OHE, United Kingdom\\
$^2$School of Physics and Astronomy, University of Nottingham, University Park, Nottingham NG7 2RD, United Kingdom\\
$^3$Department of Physics and Astronomy, University of Edinburgh, King's Buildings, Mayfield Road, Edinburgh EH9 3JZ, United Kingdom}


\begin{abstract}
We build upon our previous analytical results for the Penna model of senescence \if{({\it Phys.~Rev.~Lett.}, {\bf 89}, 288103 (2002))}\fi
to include positive mutations. We investigate whether a small but non-zero positive 
mutation rate gives qualitatively different results to the traditional Penna model in which no positive mutations 
are considered. We find that the high-lifespan tail of the distribution is radically changed in structure, but that there is not much effect on the bulk of the population. The mortality plateau that we found previously for a stochastic
generalization of the Penna model is stable to a small positive mutation rate. 

\end{abstract}

\maketitle

\section{Introduction}

The Penna model, introduced by Penna in 1995 \cite{Penna}, has been extensively studied through simulation 
\cite{Stauffer_MC,Rev_penna_oliveira}.
It models senescence in an asexually reproducing population, with the lifetime of an individual encoded in a simple bit-string model of its genome. During reproduction the bit-string is allowed to copy and mutate, so that
the population evolves a distribution of genetic lifespans. The model is powerful
enough to reproduce some aspects of observed real-life behaviour, and adaptable enough to allow modifications aimed at greater realism \cite{Salmon,Possums}. 

One shortcoming of the standard Penna model is that a mutated genome is always worse than its 
parent \cite{Penna}. While this is a reasonable assumption, given that
the rate of beneficial mutations is small compared to that of harmful ones, very little work has been carried
out on Penna models with positive mutations \cite{OliveiraPosMut}. 
Here we develop an analytical solution to a Penna model with
a small rate of positive mutation. Our analysis builds upon previous work where a solution to a generalized class of
stochastic Penna models was presented \cite{Coe_Mao_Cates}. All populations are considered in the thermodynamic 
limit of a large population where statistical fluctuations and effects of discretization can be ignored.
The exact limiting behavior found below gives insight into any population large enough for these two effects to be
unimportant.

In the standard (deterministic) Penna model, the bit-string determines the lifespan of an organism directly. One way
 (but not the only way \cite{Coe_interp}) to interpret this is to say that there is a set of hereditable diseases, 
which strike the organism at a set of fixed ages during its lifetime. Once it has developed a given number of these,
 the organism dies. Although the age of death is thus programmed from birth, an individual is in effect reading the 
bit string sequentially through its life; it drops dead after encountering the specified number of defective bits. 
(We call these 1s; the rest are 0s.) So long as it lives, the organism produces offspring at a steady rate; these 
inherit the parental genome, with a small rate $m$ of harmful mutation ($0\to 1$) and, in this paper, 
an even smaller rate $\alpha$ for positive mutation, $1\to 0$.

The simplest of all Penna models is that in which an organism dies after developing a single disease. For clarity of argument we confine our discussion to this version of the Penna model in Sections \ref{simple},\ref{positive}. 
However, the analysis of positive mutations that we present here can be adapted to any of the more 
sophisticated Penna model variants 
we have solved \cite{Analytical_solution}. In Section \ref{mortality} we address a stochastic variant showing a mortality plateau \cite{Coe_Mao_Cates}.

\section{A simple Penna model}
\label{simple}

\if{
}\fi

We consider first the single-disease simple Penna model, without positive mutation. The sites along the bit-string are numbered $x=0,1,2,....$; an organism in whose bit-string the first $1$ occurs on site $l$ will
live for exactly $l$ timesteps and thus has `genetic lifespan' $l$.
Note that an organism with $l = 0$ (i.e., a $1$ at site $x=0$ in the string) will die instantly and never contributes to the population. Note also that $x$ can be thought of as the age of an organism; so long as $x<l$, it remains alive.

The number of organisms with age $x$ and genetic lifespan $l$ at timestep $j$ is denoted by $n_j(x,l)$.
Newborn organisms with genetic lifespan $l$ can be produced either as copies of organisms with the same $l$, or
as mutated copies of organisms of lifespan $l'>l$. In both cases, sites $0...l-1$ must go unmutated. These dynamics give the following discrete evolution equation 
\begin{eqnarray}
n_{j+1}(0,l)&=&be^{-\beta l} \sum_{x=0}^\infty n_j(x,l)\label{SPM_1} \\
&+& mbe^{-\beta l}\sum_{l'>l}^\infty \sum_{x=0}^\infty n_j(x,l')\nonumber
\end{eqnarray}
where $e^{-\beta}=1-m$, with $m$ the (small) mutation rate and $b$ the birth rate.
The sum over ages of $n(x,l)$ is defined to be $n(l)$ and can be evaluated at steady state 
to give $n(l)=l\, n(0,l)$. At steady state, the size of any part of the population is time-independent.
Manipulation of (\ref{SPM_1}) then gives a recursion relation for the relative sizes of sub-populations at 
steady state:
\begin{equation}
\frac{n(l+1)}{n(l)}=\frac{l+1}{l}\frac{ e^{\beta l} - bl }
{ e^{\beta(l+1)} - b(l+1)e^{-\beta}}.\label{SPM_solution}
\end{equation}

For the population to remain finite there must be a maximum sustainable gentic lifespan $l_{\mathrm{max}}$; for $l>l_\max$, $n(l)=0$ (see Section \ref{paradox} for further discussion).
Any sub-population with a smaller genetic lifespan is partly reliant on a flux, by mutation, from longer lived
sub-populations. These two conditions give restrictions on the choice of $l_{\mathrm{max}}$, $b$ and $\beta$, as explained, and confirmed by simulation, in \cite{Coe_Mao_Cates}:
\begin{eqnarray}
l_{\mathrm{max}}&<&\frac{1}{1-e^{-\beta}}\label{SPM_Lmax},\\
b&=&\frac{1}{l_{\mathrm{max}}}e^{\beta l_{\mathrm{max}}}.\label{SPM_b}
\end{eqnarray}


\section{Solution with positive mutation}
\label{positive}
We now introduce a small positive mutation rate $\alpha$ into the simple one-disease Penna model just described.
Just as a harmful mutation converts a $0$ into a $1$,
a positive mutation rate $\alpha$ converts a $1$ into a $0$ with probability $\alpha$.
The rate $\alpha$ is taken to be sufficiently small that there is no chance of multiple 
positive mutations occuring on the same organism.
Further, we assume that $\alpha$ is small compared to the rate $\beta$ of harmful mutations. This ensures that
after the first $1$ in an organism's bit string, the remaining bits are, to high accuracy, all $1$'s due to the accumulated effects of harmful 
mutations. 
In the absence of positive mutation this is clearly the case, as the accumulation of harmful mutations 
is irreversible and there is no evolutionary pressure for an organism to have healthy sites on its bit-string beyond the site $x=l$. Positive mutations allow this accumulation to be reversed, but so long as $\alpha \ll \beta$, there remains no evolutionary pressure on sites after the first $1$; an organism's bit-string can thus be taken to consist of entirely $1$'s after the first $1$.
This is a very strong condition, and in steady state it allows (as in Section \ref{simple}) any bit-string to be characterized 
by a single number, the genetic lifespan
$l$. \if{This $l$ has exactly the same meaning as in the model of Section \ref{simple}: it is the position on the 
bit-string of the first, deadly mutation.}\fi 


\subsection{Dynamical equations}
When considering mutations on an offspring, we impose that positive mutations take place first, then negative mutations.
This is to further enforce the weak nature of positive mutations, preventing them from
overriding harmful mutations in the same time-step.

Introducing the positive mutation rate $\alpha$ then gives a slightly modified equation for $n_j(0,l)$ in place of Eq.(\ref{SPM_1}):
\begin{eqnarray}
n_{j+1}(0,l)&=&(1-\alpha)be^{-\beta l} \sum_{x=0}^\infty n_j(x,l)\\
&+& \alpha b e^{-\beta l} \sum_{x=0}^\infty n_j(x,l-1)\nonumber\\ 
&+& (1-\alpha)mbe^{-\beta l}\sum_{l'>l}^\infty \sum_{x=0}^\infty n_j(x,l')\nonumber\\
&+& \alpha mbe^{-\beta l}\sum_{l''\geq l}^\infty \sum_{x=0}^\infty n_j(x,l'').\nonumber
\end{eqnarray}  
The first term on the right corresponds to organisms with genetic lifespan $l$ reproducing with no 
harmful mutations at sites $x<l$ and no positive mutation at $x=l$. 
(Note that there are $l$ sites with $x<l$; the factor $e^{-\beta l}$ is the probability of no harmful mutation at any of these.)
The second term corresponds to offspring
from organisms with lifespan $l-1$ with one positive mutation (at site $l-1$) and no harmful mutations at sites $x<l$. 
The third term gives mutated offspring from longer lived organisms (of lifespan $l'>l$) without positive mutation, without harmful
mutation for sites $x<l$, but with harmful mutation at site $x = l$. The final term gives mutated offspring from 
longer lived organisms of lifespan $l''$ where positive mutation occurs at site $x = l''$ but is negated by harmful mutation at $l$, with no harmful mutation
for sites $x<l$. 

Defining $n(l)=\sum_x^\infty n(x,l)$ as the total number of individuals of lifespan $l$, in the steady state
we find
\begin{eqnarray}
0&=&\Bigg( (1-\alpha)be^{-\beta l} +\alpha mbe^{-\beta l} - \frac{1}{l} \Bigg)n(l)\\ 
&+&\alpha b e^{-\beta l} n(l-1) 
+ mbe^{-\beta l}\sum_{l'>l}^\infty n(l').\nonumber
\end{eqnarray}  
Writing a similar expression for $n(l+1)$ allows construction of a recursion relationship between $n(l)$, 
$n(l-1)$ and $n(l+1)$.
(Note that $n(0)$ is known to be zero.) This can be written, for given $\beta$, $b$ and $\alpha$, as:
\begin{eqnarray}
&&n(l+1)=\frac{l+1}{l}\label{srecur}\\
&&\times \frac{\big[e^{\beta l} - (1-\alpha-\alpha e^{-\beta})bl\big]n(l)-\alpha b l n(l-1)}
{e^{\beta(l+1)} -(1-\alpha)b(l+1)e^{-\beta}}.\nonumber
\end{eqnarray}
Note that if $\alpha$ is set to zero, Eq.(\ref{SPM_solution}) is recovered.

\subsection{Subtleties at large $l$}
Deducing boundary conditions at large $l$, in the presence of positive muations, is non-trivial. The possibility
that an infinitely long lived organism can evolve from a population of shorter genetic lifespans means that no part of the population is uniquely self sustaining. 
This is in contrast to the simple Penna model ($\alpha = 0$) where the
maximum sustainable genetic lifespan $l_\max$ remains finite in the thermodynamic limit of a large population; individuals of $l = l_\max$ reproduce unmutated offspring at a rate that precisely balances their own death rate. With positive mutation present this cannot be true. Organisms of putative maximal lifespan $l_\max$ can produce positively mutated offspring with 
a longer lifespan, and in a large enough population there can be no $l_\max$. 

Instead we insist that, for a steady state distribution to be physical, the population must be finite (with the thermodynamic limit taken only at the end). The effects of this requirement can be seen by taking the limit of an arbitrarily small birth rate $b$ When taking this limit we do not apply the steady
state conditions (\ref{SPM_Lmax},\ref{SPM_b}) which hold for the simple Penna model only.
In the limit of a vanishing birth rate, both the simple Penna model and Penna model with positive mutations
return a recursion relation for the relative sizes of successive sub-populations
as follows:
\begin{equation}
\frac{n(l+1)}{n(l)}=\frac{l+1}{l}e^{-\beta }.\label{sub}
\end{equation}
This limiting expression is independent of the birth rate so that, as $b$ tends to zero the population can remain 
non-zero, with a distribution that is solely dependent on $\beta$.
This non-intuitive result can be explained by the behaviour of organisms with infinite $l$. An infinitely 
long-lived organism can reproduce during its lifetime despite an arbitrarily small birth rate $b$. Moreover, in the limit of $b\to 0$, the mutated
offspring of this `super organism' make up the entire population. However, a population with even one super organism cannot be finite since, according to (\ref{sub}), $n(l\to\infty)/n(1) = 0$. It is clear, therefore, that steady state solutions of the recursion (\ref{srecur}) involving the presence of a super organism are not physical, and should be discarded.

To summarize, in the simple Penna model of Section \ref{simple} the requirement of a finite population directly imposes an $l_\max$. When positive mutations are allowed there is no $l_\max$, but the population is
still required to be finite. Thus the steady state must not contain super organisms if it is to represent a physical population, and the thermodynamic limit must be taken so as to exclude them.

\label{paradox}
\if{
}\fi

\subsection{Results}
\begin{figure}
 \includegraphics[width=3in]{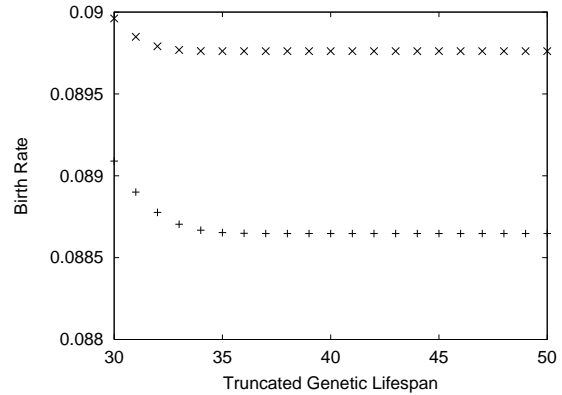}
 \caption{\label{PMfig3}
Birth rate $b$ is plotted against the trucated genetic lifepan $l_c$ for $\alpha=0.001$ ($\times$) and 
$\alpha=0.005$ ($+$).
In both cases $\beta=\frac{1}{30}$. Genetic lifespan is measured in timesteps.} 
\end{figure}

An acceptable steady state can be found by imposing an artificial maximum genetic lifespan $l_c$ beyond which $n(l)$ is 
taken to be zero.
With this artificial cut-off, for specified $\alpha$ and $\beta$, $b$ can be found so as to satisfy the steady state conditions. 
As $l_c$ approaches infinity the difference between this approximate steady state
and the real steady state will vanish for a sufficiently large population. The convergence of $b$ with increasing $l_c$ is shown in fig \ref{PMfig3}.
We require that at $l_c$ the values of $n(l_c)$ and $n(l_c-1)$
predicted by the positive mutation recursion relation satisfy to high order
\begin{equation}
\frac{n(l_c)}{n(l_c-1)}=\frac{\alpha b l_c} 
{e^{\beta l_c} -\alpha m b l_c - (1-\alpha)b l_c}.
\end{equation}
This ensures that contributions from longer lived organisms vanish, preventing the existence of a super organism
as $l_c$ tends to infinity.
This procedure resolves the paradoxes associated with the thermodynamic limit that were set out above, and allows us to find all the population properties from the recursion derived earlier.

\begin{figure}
 \includegraphics[width=3in]{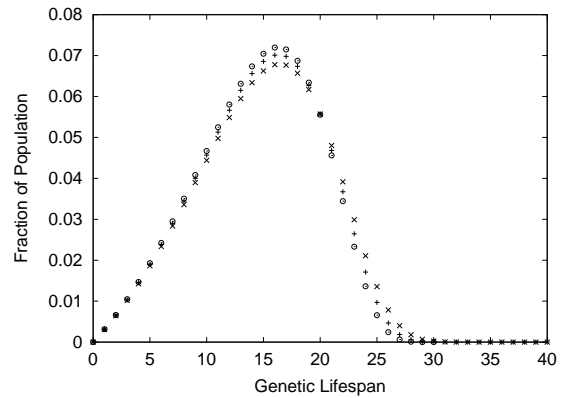}
 \caption{\label{PMfig4}Lifespan distributions for a Penna model with positive mutations in which $\alpha=0.005$
($\times$) and $\alpha=0.001$ ($+$) and without positive mutations ($\odot$). In each case $\beta=\frac{1}{30}$.
Genetic lifespan is measured in timesteps.}
\end{figure}

Analytical results, found by this method, for a Penna model with a small but non-zero $\alpha$ are shown in fig \ref{PMfig4}.
The population distributions with and without positive mutations are not dissimilar, but there are crucial differences. As discussed, with positive
mutations allowed there is no maximum genetic lifespan. What is observed instead is a small but non-zero
population beyond what would be $l_\max$ in the simple Penna model ($\alpha = 0$). The remaining distribution closely resembles $\alpha = 0$, with $l_\max$ taking its largest allowed value for the specified $\beta$.

Note that, depending on the size and history of the population, the maximum lifespan of a simple Penna population may be less than 
the largest permissible value of $l_\max$ \cite{Coe_Mao_Cates}, due to the effect of Muller's rachet \cite{Muller}. In other words, if a fluctuation within a finite population causes the self-sustaining sub-population of longest-lived individuals to die off, they can never return and the maximum lifespan is permanently reduced.
With positive mutations this loss of fitness due to statistical
fluctuations is reversible; positive mutations can act to restore the 
mean fitness of a population which has fallen below the maximum permitted by the simple Penna model.
 
\section{Effect on mortality plateau}
\label{mortality}
In \cite{Coe_Mao_Cates} we demonstrate that a Penna model with a modified survival function can exhibit a mortality
plateau at advanced ages \cite{plateau,weitz}). The strict deterministic nature of the original Penna model is relaxed; organisms
with genetic lifespan $l$ do not necessarily die at age $l$, though this is their expected lifespan. The `step function' survival up to age $l$ of the simple Penna model
is softened to a Fermi-like function whose width parameter $w$ controls the genetic indeterminacy in the age of death. 

\begin{figure}
 \includegraphics[width=3in]{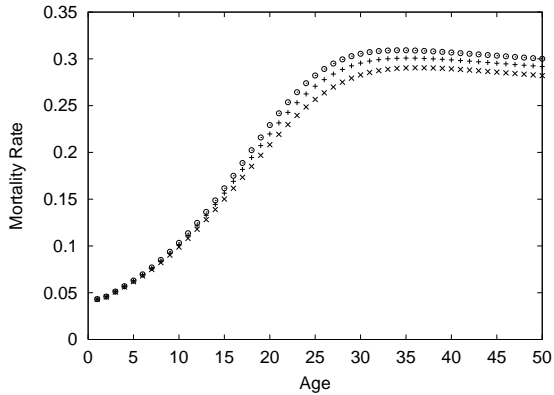}
 \caption{\label{mortPlot}Mortality rates for a Penna model with positive mutations and Fermi like survivability 
function in which $w=0.12$, $\alpha=0.005$
($\times$) and $\alpha=0.001$ ($+$) and without positive mutations ($\odot$). In each case $\beta=\frac{1}{30}$.
Age is measured in timesteps.}
\end{figure}

We observe that for a small positive mutation rate the mortality plateau is preserved (see fig \ref{mortPlot}).
This plateau is the result of organisms living beyond $l$ `on borrowed time' \cite{Coe_Mao_Cates}. Even with $\alpha > 0$,
the majority of the organisms at advanced ages are rare survivors to ages far beyond $l$; and the mortality of these survivors is almost constant. The variation in mortality rates as $\alpha$ increases comes from the increased fraction of organisms with longer genetic 
lifespans (see fig \ref{PMfig4}). Since a positive mutation rate acts to increase the average
lifespan of the population a corresponding decrease in mortality is to be expected.

\section{conclusion}

The nature of the positive mutations we have considered differs from that investigated by Oliveira  et al 
\cite{OliveiraPosMut} in that it does not increase the mean fitness of the population over time. 
Our small positive mutation rate is a relatively weak effect and while it is able to
restore the mean fitness of a population if pushed away from steady state (offering limited protection from Muller's ratchet)
it cannot improve the population's fitness in a sustained manner. Strong positive mutations,
such as those considered by Oliveira et al, are capable of sustained improvement in the fitness of 
a population but will take place over a far greater timescale than the weak positive mutations we have considered. 

Our analysis is limited to a small positive mutation rate. If the positive mutation rate 
were to become comparable to the negative one \cite{posMutPRL,posMut}(it must remain smaller for there to be a possible steady state),
then the assumption that there is no evolutionary pressure on bits after the terminal one would become invalid
and our description of the Penna string would break down. However it
has long been accepted that positive mutation is extremely rare compared to harmful mutation; as such this 
work addresses the relevant regime.

\begin{acknowledgements}
The authors would like to thank D.~Khemelnitski, M.~Rutter and J.~Trail for useful advice and discussion.
\end{acknowledgements}

\end{document}